\documentclass[twocolumn,preprintnumbers,amsmath,amssymb,superscriptaddress,prb,longbibliography]{revtex4-1}

\usepackage{graphicx,here}
\usepackage{dcolumn}
\usepackage{bm}
\usepackage{enumerate}

\newcommand{\myvec}[1]{\mbox{\boldmath $#1$}}

\begin{document}

\preprint{APS/123-QED}

\title{Polarized neutron scattering study on the centrosymmetric skyrmion host material Gd$_2$PdSi$_3$}

\author{Jiwon Ju}
\author{Hiraku Saito}
\affiliation{The Institute for Solid State Physics, the University of Tokyo, Kashiwa, Chiba, 277-8581, Japan}
\author{Takashi Kurumaji}
\affiliation{Department of Advanced Materials Science, University of Tokyo, Kashiwa 277-8561, Japan}
\author{Max Hirschberger}
\affiliation{RIKEN Center for Emergent Matter Science (CEMS), Saitama 351-0198, Japan.}
\affiliation{Department of Applied Physics and Quantum Phase Electronics Center (QPEC), University of Tokyo, Tokyo 113-8656, Japan}
\author{Akiko Kikkawa}
\author{Yasujiro Taguchi}
\affiliation{RIKEN Center for Emergent Matter Science (CEMS), Saitama 351-0198, Japan.}
\author{Taka-hisa Arima}
\affiliation{RIKEN Center for Emergent Matter Science (CEMS), Saitama 351-0198, Japan.}
\affiliation{Department of Advanced Materials Science, University of Tokyo, Kashiwa 277-8561, Japan,}
\author{Yoshinori Tokura}
\affiliation{RIKEN Center for Emergent Matter Science (CEMS), Saitama 351-0198, Japan.}
\affiliation{Department of Applied Physics and Quantum Phase Electronics Center (QPEC), University of Tokyo, Tokyo 113-8656, Japan}
\affiliation{Tokyo College, University of Tokyo, Tokyo 113-8656, Japan,}
\author{Taro Nakajima}
\email{taro.nakajima@issp.u-tokyo.ac.jp}
\affiliation{The Institute for Solid State Physics, the University of Tokyo, Kashiwa, Chiba, 277-8581, Japan}
\affiliation{RIKEN Center for Emergent Matter Science (CEMS), Saitama 351-0198, Japan.}

\begin{abstract}
We have investigated magnetic structures of the centrosymmetrric skyrmion material Gd$_2$PdSi$_3$ by means of polarized neutron scattering near zero field with an isotope-$^{160}$Gd-enriched single crystal. %
In a previous study, magnetic structures in Gd$_2$PdSi$_3$ at low temperatures were studied by resonant X-ray scattering measurements [T. Kurumaji {\it et al.} Science {\bf 365}, 914 (2019)]. %
The present polarized neutron results confirm that the magnetic structure in zero field has elliptic screw-type magnetic modulation with a propagation vector of $(q,0,0)$ with $q\sim 0.14$ and its equivalents. %
As the temperature increases, the system undergoes a magnetic phase transition while keeping the incommensurate $q$-vector of $(q,0,0)$. %
We found that the thermally-induced phase has sinusoidal magnetic modulations with magnetic moments perpendicular  both to the $c$ axis and to the $\myvec{q}$-vector. %
We also investigate the spin-helicity degree of freedom in the ground state by polarized neutrons, revealing that the system contains equal fractions of the left-handed and right-handed screw-type orders as expected from the centrosymmetric crystal structure. %
\end{abstract}


\maketitle

\section{INTRODUCTION}
Gadolinium based intermetallics have recently attracted increasing attention since a magnetic skyrmion lattice (SkL), which is a regular arrangement of topologically-nontrivial vortex-like spin objects, with a large topological Hall effect was discovered in a centrosymmetic triangular lattice magnet Gd$_2$PdSi$_3$ \cite{GPS_Kurumaji_Science}. %
Magnetic skyrmions were predicted, discovered, and have been extensively studied in noncentrosymmetric magnets with ferromagnetic exchange interactions and Dzyaloshinskii-Moriya (DM) interactions \cite{Bogdanov_1989,Sk_Theory_Nature,MnSi_Science,Yu_Nature,SkL_FeCoSi,Cu2OSeO3_Science,GaV4S8_NMat,CoZnMn_NCOMN,Kurumaji_VOSO}. %
However, the recent studies on centrosymmetric itinerant Gd compounds, such as hexagonal Gd$_2$PdSi$_3$, breathing kagom\'e Gd$_3$Ru$_4$Al$_{12}$ \cite{GRA_Max_NCom} and tetragonal GdRu$_2$Si$_2$ \cite{GRS_Khanh_NPhys}, have demonstrated that  SkL states can be realized without relying on  DM interactions. %
The microscopic mechanism to stabilize the SkL states in the centrosymmetric systems is still under debate. %
Previous studies on localized spin models pointed out the importance of geometrical spin frustration \cite{PRL_Okubo_MonteCarlo,NCom_Leonov_frustratedmag}. %
As for itinerant models, numerical calculations based on a Kondo lattice model revealed that couplings between conduction electrons and localized magnetic moments can lead to a variety of topological spin orders including SkL states \cite{PRB_Hayami_biquadratic}, assuming that the Ruderman-Kittel-Kasuya-Yosida (RKKY) interaction determines the magnetic modulation vectors. %
By contrast, a recent ab-initio calculation pointed out that orbital-dependent exchange interactions between the Gd spins are essential for the incommensurate magnetic modulations in the Gd-based intermetallics \cite{PRL_Nomoto_abinitio}. %
To examine these models, it is necessary to investigate microscopic parameters of the existing centrosymmetric skyrmion materials experimentally. %
For instance, a recent neutron scattering study on a polycrystalline sample of Gd$_2$PdSi$_3$ has suggested that the system should have long-range magnetic interactions, which supports the RKKY picture \cite{Paddison_arxiv}. %
In the present study, we performed polarized neutron scattering on a single crystal of isotope-enriched $^{160}$Gd$_2$PdSi$_3$, and investigate anisotropy and chirality of the magnetic structures in zero field. %

Gd$_2$PdSi$_3$ has an AlB$_2$-like hexagonal structure, in which triangular lattice layers of magnetic Gd atoms are  stacked along the $c$ axis. %
This system exhibits magnetic orders below 21 K \cite{PRB_Saha_MR}. %
Kurumaji \textit{et al.} performed resonant X-ray scattering measurements in the ground state, and found incommensurate magnetic reflections with magnetic modulation wave vectors ($\myvec{q}$-vectors) of $(q,0,0)$, $(0,q,0)$, and $(q,-q,0)$, where $q\sim 0.14$ \cite{GPS_Kurumaji_Science}. %
These $\myvec{q}$-vectors are equivalent to each other owing to the hexagonal symmetry of the crystal\cite{comment1}. %
By analyzing the polarization of the scattered X-rays, they concluded that the ground state has screw-type magnetic modulations. %
To further corroborate the magnetic structure in Gd$_2$PdSi$_3$, we have performed polarized neutron scattering. %
While resonant X-ray scattering is  sensitive not only to the magnetic orders but also charge density waves (CDWs), which were actually reported in a recent scanning tunneling microscopy study on GdRu$_2$Si$_2$\cite{GRS_Yasui_NCom}, polarized neutron scattering can unambiguously distinguish magnetic modulations from charge/lattice modulations. %

There are also several open questions regarding the magnetic orders in Gd$_2$PdSi$_3$. %
One is thermally-induced phase transitions in zero field. %
Recent specific heat \cite{PRB_Hirschberger_HF} and dilatometry \cite{PRB_Spachmann_dilatometry} measurements reported the existence of an intermediate phase between the paramagnetic phase and the ground state. %
We here investigate the temperature dependence of the magnetic structure by polarized neutron scattering. %
Another problem is the the sense of spin helix.  %
In centrosymmetric systems, left-handed and right-handed screw-type magnetic modulations should be energetically equivalent to each other. %
This degeneracy can be lifted when the system is noncentrosymmetric. %
In fact, previous polarized neutron scattering and transmission electron microscopy studies on an archetypal skyrmion host compounds MnSi and (Fe,Co)Si revealed one-to-one correspondences between magnetic and crystallographic chirality in single crystal samples \cite{MnSi_chirality_JPSJ,MnSi_chirality_Morikawa_PRB}.
In the present study, we also investigate the magnetic chirality in Gd$_2$PdSi$_3$ by polarized neutron scattering. %

\section{EXPERIMENTAL DETAILS}

We grew an isotope-enriched $^{160}$Gd$_2$PdSi$_3$ single crystal by the floating zone method in order to reduce the strong neutron absorption of Gd. %
The crystal was cut into a plate shape with thickness of 1.0 mm and area of 16.2 mm$^2$. %
The experiment was carried out by using the polarized neutron triple-axis spectrometer PONTA in Japan Research Reactor 3 (JRR-3). %
The sample was mounted in a closed-cycle $^4$He refrigerator with the $(H,K,0)$ scattering plane. %
The spectrometer was operated in two longitudinal polarization analysis modes, which will be explained in the following section. %
A spin-polarized incident neutron beam with the energy of 13.7 meV was obtained by a Heusler(111) crystal monochromator. %
The directions of the neutron spins were controlled by a spin flipper, guide fields, and a Helmholtz coil. %
We also employed a Heusler(111) crystal analyzer to select the energy and spin states of scattered neutrons. %
We measured spin-flip (SF) and non-spin-flip (NSF) intensities by changing the spin state of the incident neutrons using the spin flipper placed between the monochromator and the sample. %
The spin polarization of the neutron beam was 0.84, which was measured by 100 nuclear Bragg reflection of the sample. %

\begin{figure}[t]
\begin{center}
\includegraphics[clip,keepaspectratio,width=8.0cm]{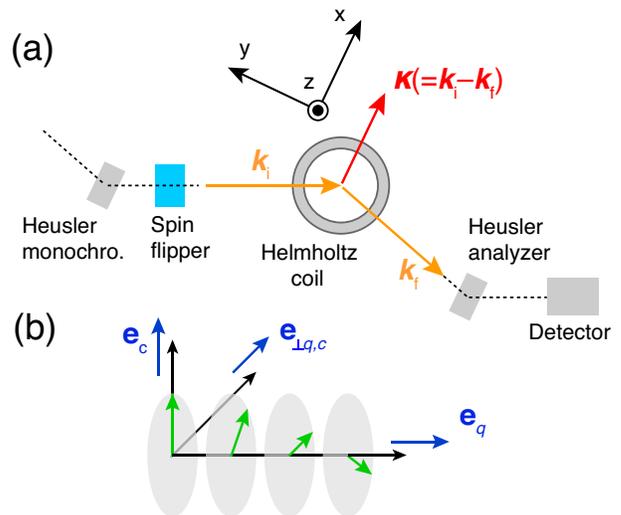}
\caption{(a) Schematic illustration showing the experimental setup and the definitions of the Cartesian coordinates $xyz$. (b) The definition of unit vectors to describe the spin components. }
\label{Defs}
\end{center}
\end{figure}

\section{DETAILS ON POLARIZED NEUTRON SCATTERING}

In neutron scattering experiments, the intensity of magnetic elastic scattering is proportional to the square of Fourier-transformed magnetic moments projected onto a plane perpendicular to the scattering vector. %
In the following, we refer to this quantity as $|\myvec{M}^{\perp}(\myvec{\kappa})|^2$, where $\myvec{\kappa} (\equiv\myvec{k}_i-\myvec{k}_f)$ is the scattering vector. %
Here, we introduce a Cartesian coordinate system $xyz$ in which the $x$ axis is parallel to $\myvec{\kappa}$, the $z$ axis is parallel to the $c$ axis, and the $y$ axis completes the right-hand set (see Fig. \ref{Defs}(a)). %
By using this coordinate system, $\myvec{M}^{\perp}(\myvec{\kappa})$ is written as $\myvec{M}^{\perp}(\myvec{\kappa})=(0,M^{\perp}_y(\myvec{\kappa}),M^{\perp}_z(\myvec{\kappa}))$. %
Note that the $x$ component should be zero, because $\myvec{M}^{\perp}(\myvec{\kappa})$ only contains the magnetic moments projected onto the $yz$ plane. %
In the longitudinal polarization analysis, intensities in SF and NSF channels are proportional to the Fourier components perpendicular and parallel to the neutron spin direction, respectively \cite{Squires}. %

In one configuration called $P_{zz}$, the direction of the incident neutron spins is parallel to the $z$ axis.
The SF and NSF intensities arising from a magnetic Bragg reflection correspond to $M^{\perp}_y(\myvec{\kappa})$ and $M^{\perp}_z(\myvec{\kappa})$, respectively. %
In addition, a lattice modulation of the same wave vector, which might be induced by a charge density wave, also contributes to the NSF intensity. %

In the other configuration termed $P_{xx}$, the neutron spin direction is turned to be parallel to the $x$ axis by a magnetic field produced by a Helmholtz coil. %
Therefore, both $M^{\perp}_y(\myvec{\kappa})$ and $M^{\perp}_z(\myvec{\kappa})$ are perpendicular to the neutron spin, and are observed in the SF channel. On the other hand, the nonmagnetic scattering is left in the NSF channel. %
Note that the guide field is so low that the magnetic structure is not affected. %

We also introduce Cartesian coordinate systems to describe the magnetic modulations corresponding to the three $\myvec{q}$-vectors. %
Figure \ref{Defs}(b) shows the three unit vectors $\myvec{e}_q$, $\myvec{e}_{\perp q,c}$ and $\myvec{e}_c$. %
We refer to amplitudes of the Fourier components with the wave vector of $\myvec{q}$ along these three axes as $m_q$, $m_{\perp q,c}$ and $m_c$. %
For example, a proper screw magnetic order has the same amplitudes of $m_{\perp q,c}$ and $m_c$, and does not have $m_q$. %
We will determine the three amplitudes from the results of the polarized neutron scattering experiment. %

\section{RESULTS AND DISCUSSIONS}

\subsection{Magnetic structure in the ground state}

\begin{figure}[t]
\begin{center}
\includegraphics[clip,keepaspectratio,width=8.5cm]{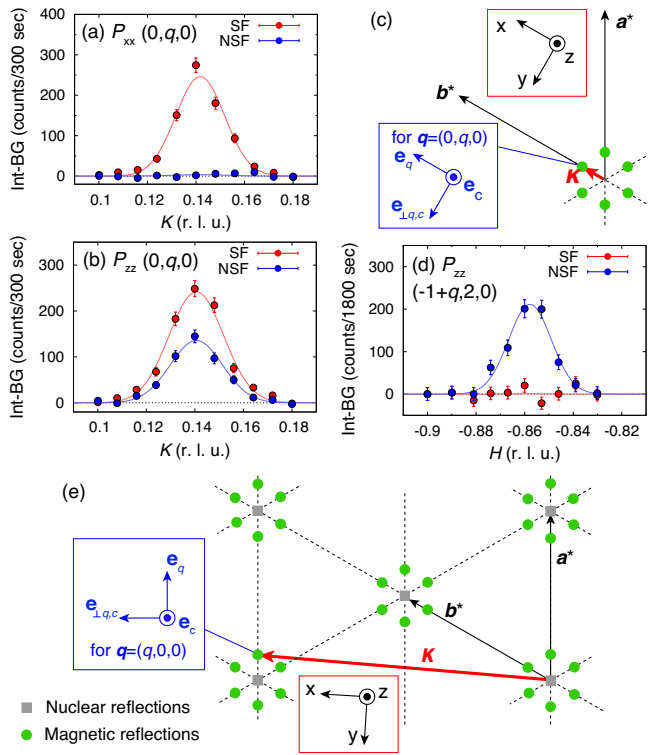}
\caption{[(a),(b)] Scattering profiles of the magnetic Bragg reflection at $(0,q,0)$ measured in the (a) $P_{xx}$ and (b) $P_{zz}$ configurations at 2.5 K in zero field. %
(c) Schematic showing the directions of $\myvec{\kappa}$ and $\myvec{q}$-vector when measuring the reflection at $(0,q,0)$. %
(d) Scattering profiles of the reflection at $(-1+q,2,0)$ measured in the $P_{zz}$ configuration at 2.5 K. %
(e) Schematic showing the directions of $\myvec{\kappa}$ and $\myvec{q}$-vector for the reflection $(-1+q,2,0)$. %
Green circles and gray squares denote positions of the magnetic and nuclear reflections, respectively. %
}
\label{Profile2K}
\end{center}
\end{figure}

The $P_{xx}$ configuration data at 2.5 K confirms the absence of non-magnetic contributions at magnetic Bragg peaks. %
Figure \ref{Profile2K}(a) shows scattering profiles of a magnetic Bragg reflection at $(0,q,0)$ measured at 2.5 K near zero field. %
We have subtracted background signals estimated from measurements at high temperatures, and  corrected the effect of the imperfect polarization. %
The reflection does not contain NSF signals, indicating that the Bragg reflection at $(0,q,0)$ is purely magnetic. %

We then measured the same reflection with the $P_{zz}$ configuration, and observed both SF and NSF intensities as shown in Fig. \ref{Profile2K}(b). %
This indicates that the magnetic structure in the ground state has both $m_c$ and $m_{\perp q,c}$ components. %
We also measured a satellite reflection at $(-1+q,2,0)$, where $\myvec{\kappa}$ is nearly perpendicular to $\myvec{e}_q$, as shown in Fig. \ref{Profile2K}(e). %
In this situation, the NSF and SF intensities correspond to $m_c$ and $m_q$ components, respectively. %
As shown in Fig. \ref{Profile2K}(d), the reflection does not contain SF scattering, indicating that the $m_q$ component is absent. %

From these results, we conclude that the ground state of this compound has a screw-type magnetic modulation. %
Note that the present measurements are not sensitive to the phase relationship between the $m_c$ and $m_{\perp q,c}$ components. %
If the incommensurate modulations of the $m_c$ and $m_{\perp q,c}$ components were in-phase, they would result in a collinear oblique-sinusoidal magnetic modulation in which magnetic moments would be pointing a direction between $\myvec{e}_c$ and $\myvec{e}_{\perp q,c}$. %
In this case, the amplitudes of each magnetic moment were sinusoidally modulated in space. %
This is not likely to happen at low temperatures, where the magnetic moments of Gd atoms are supposed to behave as classical Heisenberg spins with a fixed amplitude of 7 $\mu_B$. %

The previous resonant X-ray study by Moody \textit{et al}. suggested that Gd$_2$PdSi$_3$ has co-planer spin orders in the $ab$-plane and CDWs with the same modulation vectors as the magnetic $\myvec{q}$-vectors \cite{Moody_arxiv}. %
However, the present polarized neutron scattering results show the existence of the $c$-axis component in the magnetic modulations and absence of nonmagnetic contribution in the magnetic Bragg peak at $(0,q,0)$, which contradicts their proposal. %
Note that the present data do not exclude the possibility of concomitant CDWs in Gd$_2$PdSi$_3$. %
As exemplified in the STM study on GdRu$_2$Si$_2$ \cite{GRS_Yasui_NCom}, the close coupling between the Gd spins and conduction electrons can induce CDWs. %
In neutron scattering, CDWs are observed when the charge modulation is large enough to induce the spatial modulations of nuclei. %
Furthermore, the parasitic CDW has a modulation vector of 2$\myvec{q}$ in general. %

We always observe three pairs of magnetic satellite reflections surrounding a reciprocal lattice point. %
This means that the system exhibits either of a multi-$\myvec{q}$ magnetic order or a multi-domain state of single-$\myvec{q}$ magnetic orders. %
In the previous study, it was pointed out that the field evolution of Hall resistivity has an additional component even in the ground state, suggesting that the ground state is a multi-$\myvec{q}$ state, specifically a meron-antimeron lattice, which can have a field-induced topological Hall effect \cite{GPS_Kurumaji_Science}. %
By contrast, a recent neutron powder diffraction study suggested that a single-$\myvec{q}$ screw-type magnetic order is more likely judging from the absolute length of the magnetic moments deduced from Rietveld analysis \cite{Paddison_arxiv}. %
To unambiguously determine the magnetic structure, real-space observations of the spin textures are desired.

\begin{figure}[t]
\begin{center}
\includegraphics[clip,keepaspectratio,width=8.0cm]{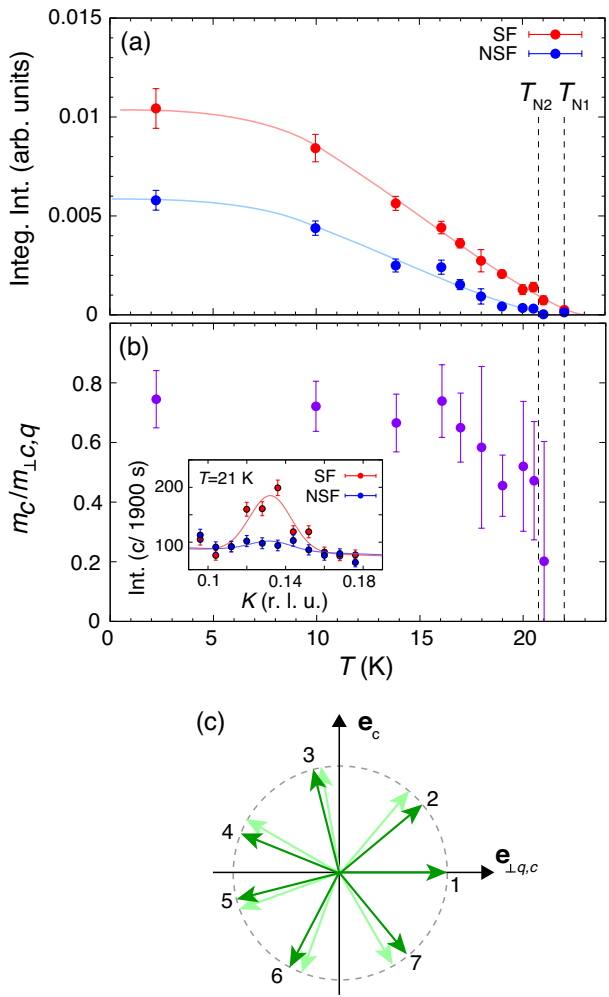}
\caption{[(a),(b)] Temperature dependence of (a) the integrated intensities of NSF and SF scatterings measured at $(0,q,0)$ and (b) the ellipticity of the screw-type magnetic order. %
Solid lines in (a) are guides to the eyes. %
Inset of (b) shows the scattering profile measured at 21 K. %
(c) Schematic illustration of the anharmonic rotation of spins in the ground state $(T<T_{\rm {N2}})$. %
The directions of the magnetic moments are projected onto the plane perpendicular to the $\myvec{q}$-vector. %
Dark green arrows show the spin directions taking into account the ellipticity of 0.7. %
The numbers correspond to relative positions of the spins along the $\myvec{q}$-vector. %
Light green arrows show the spin directions without ellipticity. %
}
\label{Tdep}
\end{center}
\end{figure}

\subsection{Temperature dependence of the magnetic structure}

The temperature dependence of the SF and NSF intensities for the magnetic Bragg reflection at $(0,q,0)$ is shown in Fig. \ref{Tdep}(a). %
Both the SF and NSF intensities monotonically decrease with increasing temperature. %
However, the NSF intensity decreases more rapidly than the SF intensity. %
In Fig. \ref{Tdep}(b), we show temperature dependence of the ellipticity, $m_c/m_{\perp q,c}$. %
At around 21 K, the NSF intensity becomes nearly zero, while the SF intensity still remains, as shown in the inset of Fig. \ref{Tdep}(b). %
By further increasing the temperature, the SF intensity also disappears above 22 K. %
These thermally-induced successive magnetic phase transitions are consistent with the previous specific heat \cite{PRB_Hirschberger_HF} and dilatometry \cite{PRB_Spachmann_dilatometry} measurements. %

The suppression of the NSF intensity shows that the intermediate phase has sinusoidal magnetic modulations with the $m_{\perp q,c}$ component. %
Although the lengths of the local magnetic moments should not change with temperature, their thermal average can be reduced and modulated as the thermal fluctuations become stronger near the critical temperature. %
These kinds of helical-to-sinusoidal magnetic phase transitions are also seen in other Gd based skyrmion host materials \cite{GRS_Khanh_NPhys,GRA_Max_NCom,AdvSci_Khanh_zoology} and in other frustrated magnets. %

We also note that the ellipticity is smaller than unity ($m_c/m_{\perp q,c}\sim 0.75$) even at the lowest temperature, where the thermal fluctuations are negligible. %
This observation is consistent with the previous resonant x-ary scattering study \cite{GPS_Kurumaji_Science}. %
The smaller Fourier component along the $c$ axis implies distortion of the screw-type structure, specifically anharmonic rotation of the magnetic moments as illustrated in Fig. \ref{Tdep}(c). %
The anharmonicity should lead to odd higher harmonic reflections. %
Assuming that all the magnetic moments have the same magnitude, we estimate that an absolute square of the Fourier amplitude for the third harmonics will be approximately two orders of magnitude smaller than that of the primary modulation. %
In the present study, such a weak higher harmonic reflections can not be detected due to a lack of statistics. %

A possible interpretation of this distortion of the magnetic structure is that the system has an easy-plane-type magnetic anisotropy. %
In most of the theoretical studies on SkLs, an easy-plane type anisotropy was considered to be unfavorable for stabilizing SkL states, and thus an easy-axis type anisotropy was imposed \cite{PRB_Hayami_anisotropy,PRL_Wang_2DEG}. %
It can be qualitatively understood that the easy-plane type anisotropy tends to support transverse conical magnetic modulations in a magnetic field parallel to the $c$ axis rather than the SkL that should possess some of the magnetic moments to point downward. %
The easy-plane magnetic anisotropy causing the partial suppression of the $c$-axis Fourier components in zero field as observed in the present study seems to be in conflict with the  above simple theoretical scenario to favor the SkL formation; this implies that additional anisotropy term and/or anisotropic exchange interactions should be taken into account to explain the magnetic structures in Gd$_2$PdSi$_3$, including the SkL state. %
For instance, a recent Monte Carlo simulation including magnetic dipole-dipole interactions reproduces a helical ground state and a field-induced SkL state \cite{Paddison_arxiv}. %
The role of the anisotropic exchange interactions was also studied in a previous study on Gd$_3$Ru$_4$Al$_{12}$, in which a combination of the easy-plane type anisotropy and anisotropic exchange interaction is necessary to  explain the observed field-angle dependence of the magnetic phase diagram \cite{NJP_Hirschberger_2021}. %

\begin{figure}[t]
\begin{center}
\includegraphics[clip,keepaspectratio,width=8.5cm]{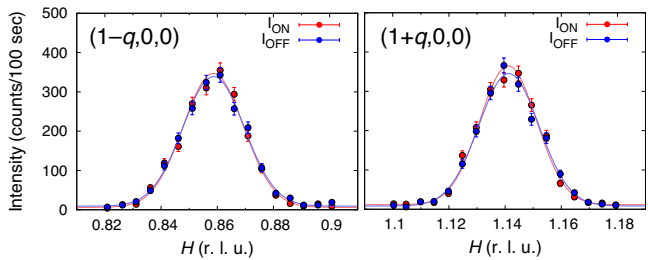}
\caption{Scattering profiles of magnetic Bragg reflections at $(1-q,0,0)$ and $(1+q,0,0)$ measured at 2.5 K. %
The spectrometer was operated in the half-polarized mode. %
The direction of the incident neutron polarization was set to be parallel or antiparallel to the scattering vector $\mbox{\boldmath $\kappa$}$. %
The data labeled "$I_{\rm{OFF}}$" and "$I_{\rm {ON}}$" were measured when the spin flipper was off and on, and correspond to the parallel and antiparallel cases, respectively. }
\label{chiral}
\end{center}
\end{figure}

\subsection{Chirality of the magnetic structure}

In the $P_{xx}$ configuration, intensities of magnetic Bragg reflections from a screw-type magnetic modulation depend on the relationship between the directions of the $\myvec{q}$-vector, the scattering vector and the neutron polarization. %
In addition the sense of the spin helix can account for asymmetry in intensity between a pair of magnetic satellite reflections. %
When the scattering vector is parallel to the screw axis of the magnetic structure, the scattering cross section for a magnetic Bragg reflection at $\myvec{\tau}\pm\myvec{q}$, where $\myvec{\tau}$ is a reciprocal lattice point, is given by %
\begin{eqnarray}
\bigg(\frac{d\sigma}{d\Omega}\bigg)_{\mbox{\boldmath $\tau$}\pm\mbox{\boldmath $q$}} %
\propto & \{(m_{c}^2+m_{\perp q,c}^2)(V^{\rm RH}+V^{\rm LH})\nonumber\\ %
  \mp& 2m_{c}m_{\perp q,c}(\mbox{\boldmath $p$}_{\rm N}\cdot\mbox{\boldmath $\hat{\kappa}$})(V^{\rm RH}-V^{\rm LH})\}.%
\label{BlumeEq}
\end{eqnarray}
 Here, $V^{\rm RH}$ and $V^{\rm LH}$ are the volume fractions with right-handed and left-handed helical magnetic orders, respectively. %
$\mbox{\boldmath $\hat{\kappa}$}$ and $\mbox{\boldmath $p$}_{\rm N}$ are unit vectors of $\mbox{\boldmath $\kappa$}$  and the polarization direction of the incident neutrons, respectively. %
Thus the scalar product  $\mbox{\boldmath $p$}_{\rm N}\cdot\mbox{\boldmath $\hat{\kappa}$}$ takes a value of either $+1$ or $-1$ depending on the neutron spin state controlled by the flipper. %
Accordingly, the intensities of the Bragg reflections also change with the direction of the neutron spins when $V^{\rm RH}$ and $V^{\rm LH}$ are not equal to each other. %

We measured magnetic Bragg reflections at $(1-q,0,0)$ and $(1+q,0,0)$ at 2.5 K in zero field. %
$\mbox{\boldmath $p$}_{\rm N}$ was set to be parallel to $\mbox{\boldmath $\kappa$}$, which was also parallel to the screw axis of the magnetic modulation with $\mbox{\boldmath $q$}=(q,0,0)$. %
We also removed the analyzer in order to measure SF intensities for both the spin-up and spin-down incident neutrons. %
Note that the absence of NSF intensities was confirmed by the data shown in Fig. \ref{Profile2K}(a). %
As shown in Fig. \ref{chiral}, the intensities did not depend on the direction of $\mbox{\boldmath $p$}_{\rm N}$ for both reflections, indicating that $V^{\rm RH}$ and $V^{\rm LH}$ are comparable to each other. %
This means that left-handed and right-handed screw-type magnetic domains are nucleated with equal volume fraction upon the magnetic phase transition to the elliptic screw-type ground state in zero field, as expected from the centrosymmetric nature of the crystal. %

\section{Summary}
We have studied magnetic structures in Gd$_2$PdSi$_3$ by means of polarized neutron scattering near zero field. %
The satellite reflections observed in the present neutron experiment are of purely magnetic origin. %
The ground state has elliptic (distorted) screw-type magnetic modulations with the smaller modulation amplitude along the $c$ axis. %
The spin component along the $c$ axis decreases with increasing temperature, and disappears at a slightly lower temperature than the transition to the paramagnetic phase. %
This indicates the existence of an intermediate phase with the sinusoidal magnetic modulations. %
These observations imply easy-plane type magnetic anisotropy in this system. %
We have also confirmed that the volume fractions of left-handed and right-handed screw-type modulations are equal to each other. %
A recent study on the itinerate helimagnet MnP demonstrated that simultaneous applications of electric current and magnetic field can control the spin helicity of the magnetic structure \cite{MnP_Onose_NCom}. %
Controlling the magnetic chirality in centrosymmetric skyrmion hosts using this method will be an interesting challenge for the future. %

\section*{Acknowledgements}

This work was partly supported by Grants-in-Aid for Scientific Research (Grants Nos. 24224009, 17K18355, 22H04463, and 21K13877) from JSPS and CREST (Grant Nos. JPMJCR20T1 and JPMJCR1874) from JST. %
The neutron scattering experiment at JRR-3 was carried out along the proposals Nos. 21507, 21401 and 22401 and partly supported by ISSP of the University of Tokyo. %
M.H. was supported by the Mizuho Foundation for the Promotion of Sciences, and by the Iketani Foundation. %

\bibliography{main}

\end{document}